\documentclass[preprints,article,accept,moreauthors]{Definitions/mdpi} 

\usepackage{epsfig, rotating, longtable, lscape, longfigure}

\firstpage{1} 
\pubvolume{1}
\issuenum{1}
\articlenumber{0}
\pubyear{2022}
\copyrightyear{2022}
\datereceived{} 
\dateaccepted{} 
\datepublished{} 
\hreflink{https://doi.org/} 

\Title{Ultraviolet and X-ray light-curves of novae observed by the {\em Neil Gehrels Swift Observatory}.}

\TitleCitation{Ultraviolet and X-ray light-curves of novae observed by the {\em Neil Gehrels Swift Observatory}}

\Author{Kim L. Page $^{1}$*, N. Paul M. Kuin$^{2}$ and Julian P. Osborne$^{1}$}

\AuthorNames{Kim Page, Paul Kuin and Julian Osborne}

\AuthorCitation{Page, K.L.; Kuin, N.P.M.; Osborne, J.P.}

\address{%
$^{1}$ \quad School of Physics \& Astronomy, University of Leicester, Leicester, Leicestershire, LE1 7RH, UK\\
$^{2}$ \quad Mullard Space Science Laboratory, University College London, Holmbury St. Mary, Dorking, Surrey, RH5 6NT, UK}

\corres{klp5@leicester.ac.uk}

\abstract{With rapid response capabilities, and a daily planning of its observing schedule, the {\em Neil Gehrels Swift Observatory} is ideal for monitoring transient and variable sources. Here we present a sample of the 12 novae with the most detailed ultraviolet (UV) follow-up by {\em Swift} -- the first uniform analysis of such UV light-curves. The fading of these specific light-curves can be modelled as power-law decays (plotting magnitude against log time), showing that the same physical processes dominate the UV emission for extended time intervals in individual objects. After the end of the nuclear burning interval, the X-ray emission drops significantly, fading by a factor of around 10--100. The UV changes, however, are of a lower amplitude, declining by 1--2 mag over the same time period. The UV light-curves typically show a break from flatter to steeper around the time at which the X-ray light-curve starts a steady decline from maximum, $\sim$~0.7–1.3~T$_{\rm SSSend}$.  Considering populations of both classical and recurrent novae, and those with main sequence or giant companions, we do not find any strong differences in the UV light-curves or their evolution, although the long-period recurrent novae are more luminous than the majority of the classical novae.}


\begin{document}


\section{Introduction}

The {\em Neil Gehrels Swift Observatory} ({\em Swift} hereafter; \citep{swift}) was launched on 2004 November 20, with the primary goal of discovering and localising Gamma-Ray Bursts (GRBs) to arcsec accuracy within a few hundred seconds, and then following the X-ray and optical afterglows as they fade. While GRBs are still an important part of {\em Swift}'s work, over the years the satellite has become the go-to mission for observations of all kinds of transients as they vary in the optical, ultraviolet (UV), X-ray and $\gamma$-ray regimes. In this paper, we focus on novae observed in the UV by the UV/Optical Telescope (UVOT; \cite{roming05}). 

Novae are thermonuclear explosions formed in accreting white dwarf (WD) systems. Material is transferred from the secondary, or companion, star onto the surface of the WD primary, until the pressure and temperature are sufficient to trigger a thermonuclear runaway (see \cite{bodeevans08, woudtribeiro14} for review articles), flinging material outwards, and shrouding the WD surface from view. At this point, a new optical source -- the nova -- is formed, with the peak optical brightness occurring when the photosphere reaches maximum expansion\footnote{While most novae are first discovered at optical wavelengths, V959~Mon was initially detected in $\gamma$-rays by the Large Area Telescope onboard {\it Fermi}, when too close to the Sun for ground-based telescopes to observe \cite{cheung12, cheung12b}. Models of nova eruptions also predict there should be a brief, soft X-ray flash after hydrogen ignition, before the optical emission is detectable. This was finally observed for the first time in 2020 by {\em eROSITA} (extended Roentgen Survey with an Imaging Telescope Array) for the nova YZ~Ret \cite{konig22}.}.
As the ejecta expand, they become optically thin, typically allowing the surface nuclear burning (and hence soft X-rays) to become visible; this is the so-called super-soft source (SSS) phase\footnote{It is, however, possible for the nuclear burning to cease before the ejecta fully clear, in which case the soft X-rays may fade away before they can be detected. In the case of V745~Sco \cite{page15}, it was suggested that only the cooling, tail-end of the SSS emission was seen, with the actual hydrogen burning having ended very quickly, placing it close to this `unobservable region'.}. Once the hydrogen shell has been burnt, the system fades back to quiescence.

There are two main groups of novae, termed Classical Novae (CNe) and Recurrent Novae (RNe), respectively. 
CNe -- which make up the majority -- have only been seen in eruption once. These systems have orbital periods of a few hours, and the secondary star is typically a late-type main sequence (MS) object \citep[e.g.,][]{darn12}.
RNe, on the other hand, have had multiple detected eruptions, with recurrence timescales of up to around 100~yr (though this is, of course, something of a selection effect, dependent on historical records; even the so-called CNe are expected to erupt more than once, with duty cycles of typically thousands of years or longer \cite[e.g.,][]{truran86}). The orbital periods of RNe cover a wide range, from a few hours to a year or more \cite{brad10, anupama08}, with the longer periods corresponding to systems where the secondary star is an evolved giant; the short period systems are more similar to CNe, with an MS or subgiant companion star.

A binary system with a higher-mass WD and higher accretion rate is expected to have a shorter recurrence time \cite{starrfield89} (though T~Pyx is unusual, in that it is thought to have a comparatively low mass WD; \cite[e.g.,][]{toff13, nelson14, chomiuk14}). Given the shorter timescales involved, RNe accrete, and then eject, less material during each nova cycle \citep[e.g.,][]{wolf13}, which means the SSS phase tends to become visible more rapidly in these systems. The optical/UV light-curves in some RNe have been noted to show a flat, plateau phase, which is speculated to arise from the re-radiation of the SSS emission from an accretion disc \cite{brad10,hachisu06,hachisu08}.
There are ten confirmed Galactic RNe \citep{brad10}, with more known in other galaxies, such as M31 \citep{shafter15, darn21}. 

Our sample contains some novae (classical and recurrent) which occurred in symbiotic binaries.
These are systems where the secondary star is a late-type giant, with outflowing wind encompassing the WD; they are sometimes known as embedded novae, to differentiate them from the so-called `symbiotic novae', which are a different population of {\em slowly-evolving} eruptive variable stars \cite{mur94,mik07,mik08}, and are not considered here.

The first nova which {\em Swift} monitored in detail was the 2006 eruption of the RN RS~Oph \cite{bode06, osborne11}. This impressive dataset showed that {\em Swift} was well suited for monitoring such sources, and this subsequently led to many more novae being observed in detail with the observatory. Examples within our own Galaxy include V458~Vul \citep{v458vul, greg11}, V2491~Cyg \citep{ibarra09, page10}, HV~Cet \citep{hvcet}, V5668~Sgr \cite{gehrz18}, V745~Sco \citep{page15}, V407~Lup \citep{aydi18} and V3890~Sgr \citep{page20a}.

The best monitored {\em Swift} light-curves up until the end of 2017 were collated by \cite{page20b}, where both X-ray and UV results were presented. In this paper, we focus in more detail on the evolution of the UV emission in a sample of 12 novae. These were chosen to be the sources best monitored by {\em Swift}-UVOT over an extended period of time, in at least one UV filter, where the light-curves follow a series of power-law decays.

While this paper considers only novae, this approach to fitting UV light-curves could be extended to other transients, such as supernovae or active galactic nuclei.

\section{Observations}

The {\em Swift}-UVOT is one of two narrow field instruments onboard {\em Swift}; the X-ray Telescope (XRT; \cite{burrows05}) is the other. The UVOT has, among other observing capabilities, six broadband filters: three optical and three UV. Given that there are a large number of ground-based observers who frequently follow novae at optical wavelengths, {\em Swift} typically utilises one or more of the UV filters when monitoring novae: $uvw1$ (central wavelength, $\lambda_c$~=~2600 \AA; Full Width at Half Maximum, FWMH~=~693~\AA), $uvm2$ ($\lambda_c$~=~2246 \AA; FWHM~=~498~\AA) and $uvw2$ ($\lambda_c$~=~1928 \AA; FWHM~=~657~\AA).  More details on the instrument can be found in \cite{poole08, breeveld11}.

Novae can peak at very bright magnitudes, some even reaching naked eye visibility -- RS~Oph, for example, peaks around magnitude 4.5 in the visual band. While this is good news for the typical stargazer, sources brighter than around magnitude 10--11 lead to significant coincidence loss in the UVOT, which is not recovered when using the standard {\em Swift} photometric tools. A technique was developed by \cite{readout} whereby the read-out streak formed by bright sources can be used to obtain measurements up to 2.4 mag brighter than the previous limit, thus extending the useful dynamic range of the instrument. We have utilised this method where relevant in the following analysis.

In 2020 September, a more accurate UVOT calibration file was released, to account for the loss of sensitivity over time. The light-curves presented here have been corrected for this degradation\footnote{See details at \url{https://www.swift.ac.uk/analysis/uvot/}.}, leading to small differences between these results and those previously published for the older novae (up to $\sim$~0.02 mag for the sources in this sample).

\begin{figure}[H]
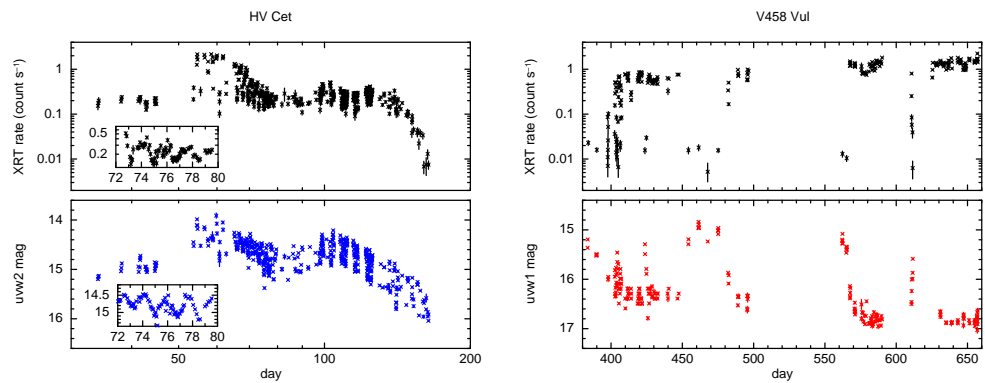

\includegraphics[width=5.0 cm,angle=-90]{HVCet_inset_XRT.eps}
\includegraphics[width=5.0 cm,angle=-90]{V458Vul_XRT.eps}
\caption{Examples of light-curves where the UV emission is not simply a series of power-law decays. Left: HV Cet. This nova shows a strong periodic oscillation in both the UV and X-ray bands, as well as an underlying slower modulation. The inset shows a zoom-in of the 1.77-d period, plotted in linear time. Right: V458 Vul. This nova shows aperiodic UV modulation, which is approximately anti-correlated with the corresponding X-ray emission. In both cases the XRT count rate is over 0.3--10~keV. \label{fig:diff}}
\end{figure}

Not only do novae show intrinsic variability as the source itself evolves, novae can be very different from each other, showing a variety of light-curve shapes \citep[e.g., ][]{page20b} across all wavebands. Some show periodic modulations (e.g., HV~Cet; \cite{hvcet}), while others brighten and fade apparently randomly, sometimes with the UV in antiphase with the X-ray emission (e.g., V458~Vul; \cite{v458vul, greg11}); see Fig.~\ref{fig:diff}. Many, however, do just (mainly) fade in the UV after peak brightness, and, in this paper, we concentrate on these novae. Specifically, we present a sample where detailed UVOT monitoring, starting promptly (within a few days) after the eruption\footnote{V959~Mon is the exception, with observations starting some time after the actual nova eruption, because of its location with respect to the Sun at outburst (see footnote 1).}, was performed in at least one UV filter, over an extended period of time. Table~\ref{tab:novae} lists the names of the novae, together with their eruption dates, orbital periods, interstellar reddening E(B$-$V), and (approximate) distances\footnote{Over the years, distance estimates to novae have changed. We use the most recent derivations by \cite{brad22a}, using parallax data from the third {\em Gaia} data release (DR3) among other methods, for all our sample bar V407~Cyg, which was not included in this catalogue.}. 
Those novae which occurred in symbiotic systems are noted in the sixth column. 

\begin{table}[H] 
\caption{Names, eruption details, orbital periods, interstellar reddening E(B$-$V), and distances of the novae in this paper, ordered by eruption date; symbiotic systems are marked in the sixth column. 
The novae marked in bold are recurrent; the eruption date given is that for the data included in this article. \\$^a$ Date of $\gamma$-ray discovery by {\em Fermi}-LAT. \\$^b$ While \citep{joshi15} refer to V1534~Sco as a symbiotic system, \cite{munari17} note that there are some inconsistencies with the evidence for a cool giant. We will therefore not treat V1534~Sco as a symbiotic in this work.\\ $^c$ There has been significant uncertainty in the distance to RS~Oph over the years, with 1.6~kpc being assumed for a long time, although it now appears this larger estimate is preferred; see discussions in \cite{brad22a, marina22} for example.  \\$^d$ https://www.aavso.org/aavso-alert-notice-752   \label{tab:novae}}
\begin{adjustwidth}{-\extralength}{0cm}
\begin{tabularx}{530pt}{CCXCCCCC}
\toprule
\textbf{Nova}	& \textbf{Alternative name} & \textbf{Eruption date UT}	& \textbf{P$_{\rm orb}$} & E(B$-$V) & \textbf{Symbiotic?} & distance (kpc) & \textbf{References}\\
\midrule
V2491 Cyg & Nova Cyg 2008 No. 2  & 2008-04-10.73 & 2.3 h & 0.23 & & 4.8 & \cite{nakano08,baklanov08,helton08,rudy08,brad22a}\\ 
{\bf U Sco}  & --  & 2010-01-28.44  & 1.23 d & 0.20 & & 6.3 & \cite{brad10,brad10b, ash15,brad22a}\\ 
V407 Cyg  & --  & 2010-03-10.80  & 43 y & 0.5 & Y & 3.9 & \cite{nishiyama10,mun90,hk19,shore11a}\\
{\bf T Pyx} & -- & 2011-04-14.24  & 1.83 h & 0.25 & & 3.6 & \cite{brad10,waagan11,shore11b,brad22a}\\
V959 Mon  & Nova Mon 2012 &  2012-06-22$^{a}$ & 7.1 h & 0.38 & & 2.9 &  \cite{fujikawa12,cheung12,cheung12b,page13,shore13,brad22a}\\
V339 Del & Nova Del 2012 & 2013-08-14.58 & 0.163 d & 0.18 & & 1.6 & \cite{nakano13,shore16,brad22a}\\
{\bf V745 Sco}   & --  & 2014-02-06.69  & 2440 d & 1.00 & Y & 8.0 & \cite{brad10,waagen14, banerjee14,brad22a}\\
V1534 Sco  & Nova Sco 2014 & 2014-03-26.85& 520 d  & 1.11& ?$^b$ & 8.2 & \cite{nishiyama14,joshi15,munari17,hk19,brad22a,brad22b}\\
V1535 Sco & Nova Sco 2015 & 2015-02-11.84& 50 d & 0.80 & Y &7.8 & \cite{nakano15, walter15, nelson15,linford17,hk19,brad22a,brad22b}\\ 
{\bf V3890 Sgr} & -- & 2019-08-27.87 & 747.6 d & 0.59 & Y & 8.5 & \cite{brad10,pereira19,mun19, evans22,brad22a,mik21}\\
V1674 Her & Nova Her 2021 & 2021-06-12.19 & 0.153 d & 0.50 & & 3.2 & \cite{patterson21,drake21,woodward21,brad22a}\\
{\bf RS Oph} & -- & 2021-08-09.54 & 453.6 d & 0.73 & Y  & 2.7$^c$ & AAVSO$^d$; \cite{brandi09,brad10,page22,brad22a,snijders87}  \\ 
\bottomrule
\end{tabularx}
\end{adjustwidth}
\end{table}
\unskip

\section{Analysis and results}

Following \cite{page13, page15, page20a}, we parameterise the UV decay as a series of power-laws. That is, we consider the magnitudes to be proportional to log(time), which is equivalent to flux proportional to time: f~$\propto$~(t/1~day)$^{-\alpha}$; T$_0$ is defined as the start of the nova eruption.
While optical/UV light-curves are traditionally plotted in terms of magnitude versus linear time, a log scale allows us to infer useful information, with trends easily visible to the eye.
A series of power-laws was thus fitted to each UVOT light-curve, using a $\chi^2$ minimisation routine for both slopes and break times; breaks were included if they were significant at the 3$\sigma$ level. 

The power-law slope $\alpha$ was then calculated as 

\begin{equation}
    \alpha = \frac{1}{2.5} . \frac{m_2 - m_1}{log\frac{t_2}{t_1}}
\label{eqn}
\end{equation}

where m$_1$ and m$_2$ are the magnitudes from the fitted model at the break times of t$_1$ and t$_2$.
For each nova, the different UV filter data were fitted independently, and all breaks are assumed to be instantaneous, with no smoothing applied.

\begin{figure}[H]
\includegraphics[width=6.3 cm,angle=-90]{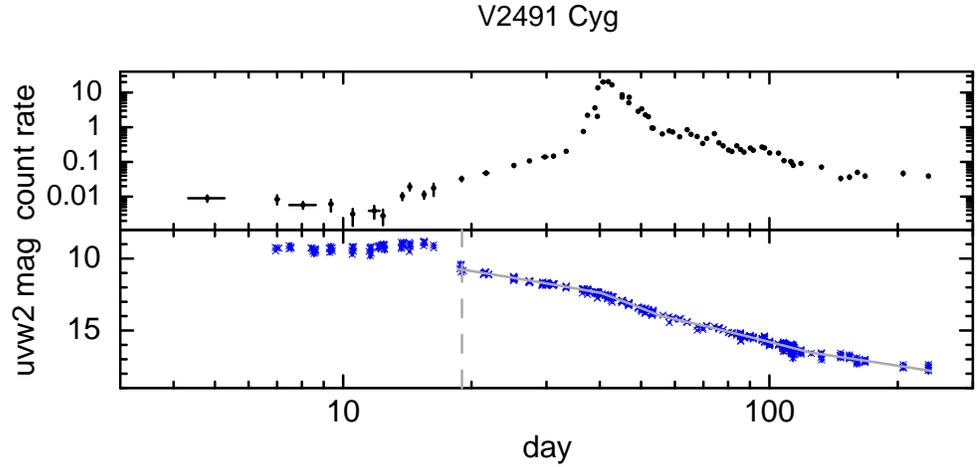}
\caption{The XRT and UVOT light-curve of V2491 Cyg, with the best-fit broken power-law decline shown as the grey solid line. The UV light-curve shows a rebrightening `cusp' after day 12, so only the data after day 19 (marked with the vertical dashed line) were fitted. The X-ray data (top) are in count~s$^{-1}$ over 0.3--10~keV.  \label{fig:V2491Cyg}}
\end{figure}

\begin{figure}[H]
\includegraphics[width=8.5 cm,angle=-90]{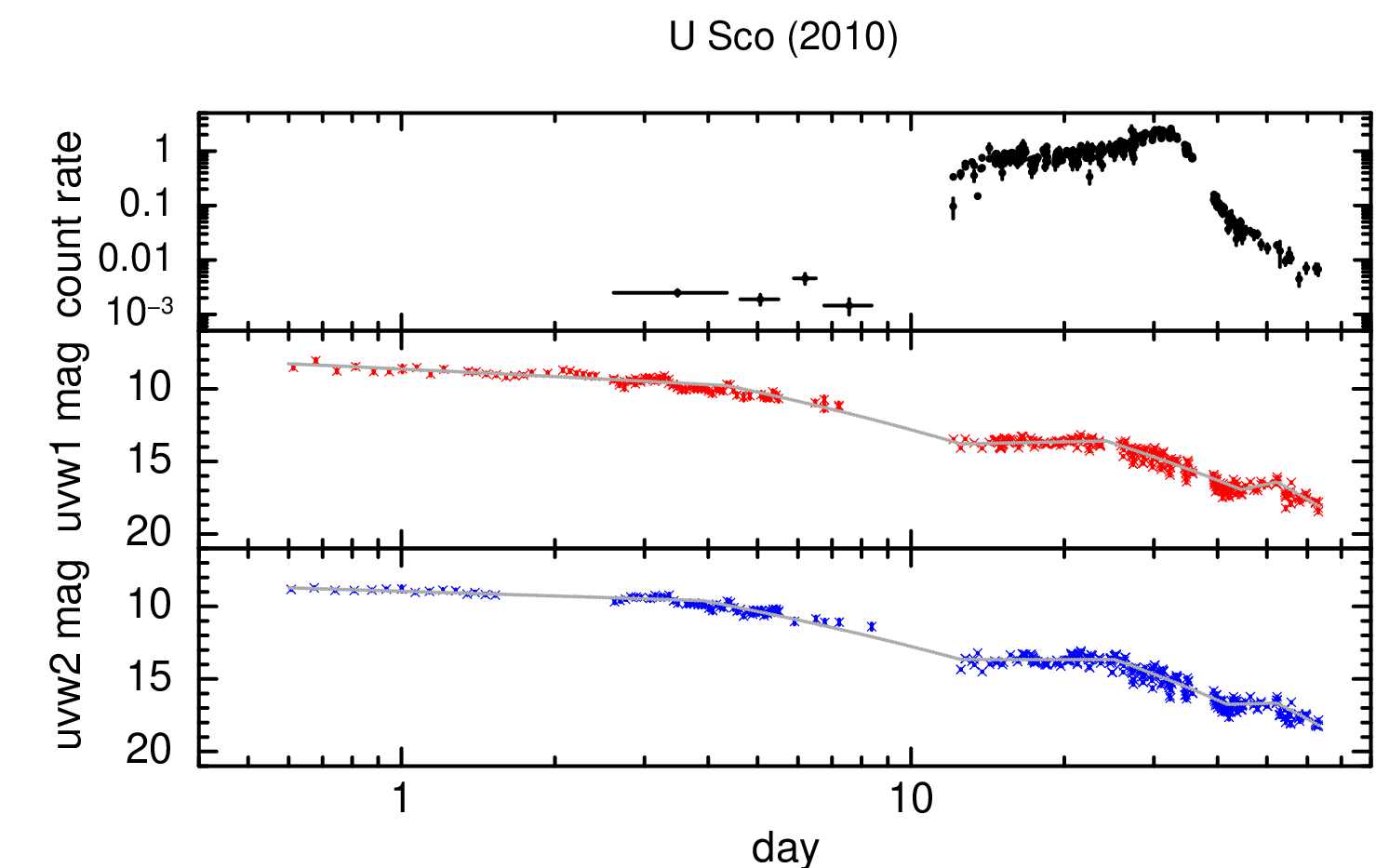}
\caption{The XRT and UVOT light-curves of U Sco, with the best-fit broken power-law declines shown as the grey solid lines. The X-ray data (top) are in count~s$^{-1}$ over 0.3--10~keV. \label{fig:USco2010}}
\end{figure}

\begin{figure}[H]
\includegraphics[width=6.3 cm,angle=-90]{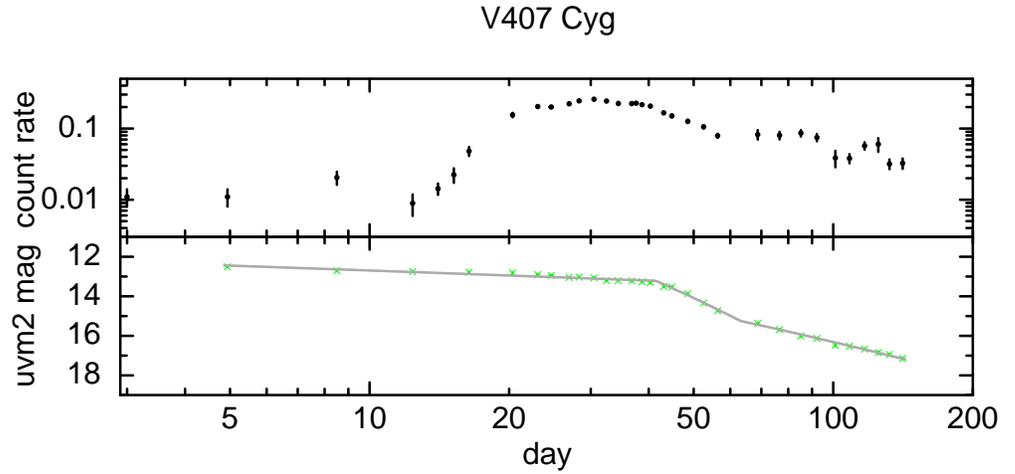}
\caption{The XRT and UVOT light-curves of V407 Cyg, with the best-fit broken power-law decline shown as the grey solid line. The X-ray data (top) are in count~s$^{-1}$ over 0.3--10~keV. \label{fig:V407Cyg}}
\end{figure}

\begin{figure}[H]
\includegraphics[width=6.3 cm,angle=-90]{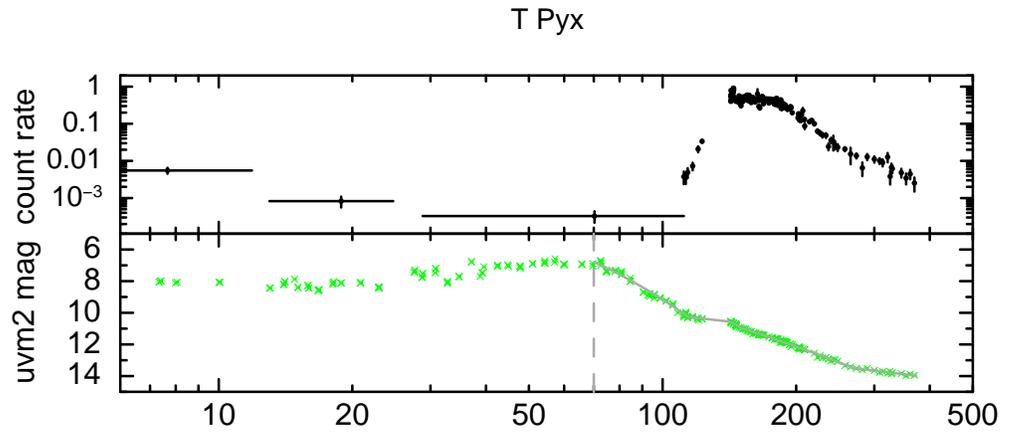}
\caption{The XRT and UVOT light-curves of T Pyx, with the best-fit broken power-law decline shown as the grey solid line. Only data after the UV peak on day 70 (marked with the vertical dashed line) were considered. The observations of T~Pyx are plotted as days since discovery; the optical peak occurred around 28 days later. The X-ray data (top) are in count~s$^{-1}$ over 0.3--10~keV. \label{fig:TPyx}}
\end{figure}

\begin{figure}[H]
\includegraphics[width=10.5 cm,angle=-90]{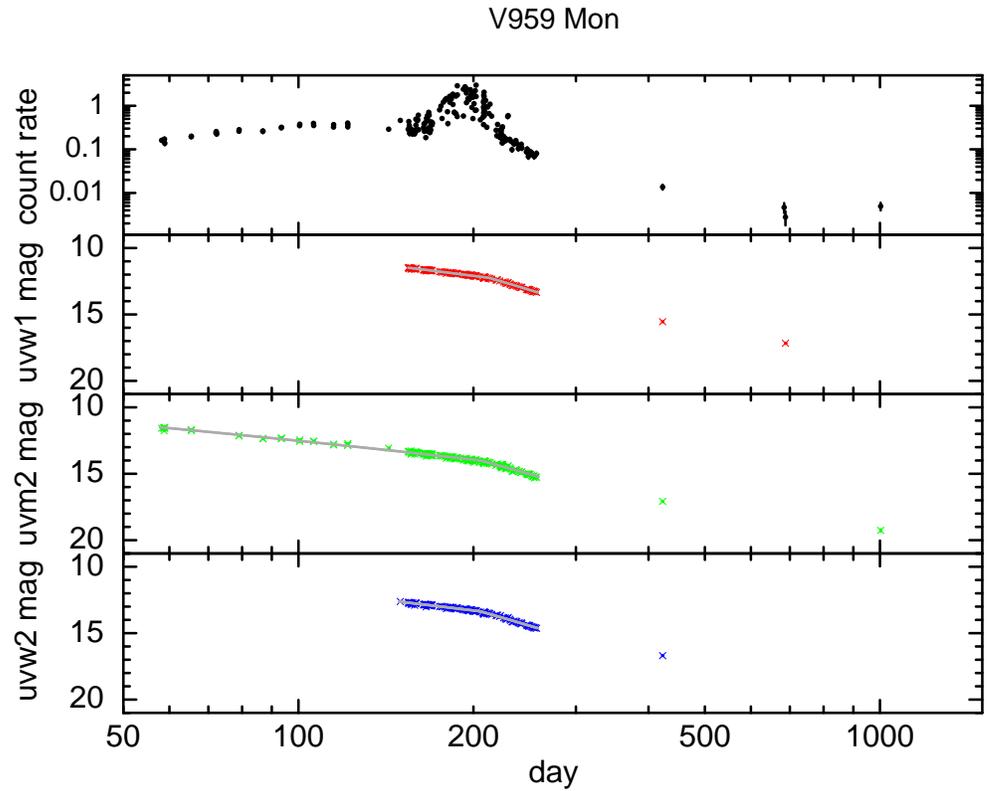}
\caption{The XRT and UVOT light-curves of V959 Mon, with the best-fit broken power-law declines shown as the grey solid lines. Late-time data beyond day 300 were excluded from the fits, because there are not enough bins to constrain the break times accurately. The observations of V959~Mon are plotted as days since {\em Fermi} discovery. The X-ray data (top) are in count~s$^{-1}$ over 0.3--10~keV. \label{fig:V959Mon}}
\end{figure}

\begin{figure}[H]
\includegraphics[width=10.5 cm,angle=-90]{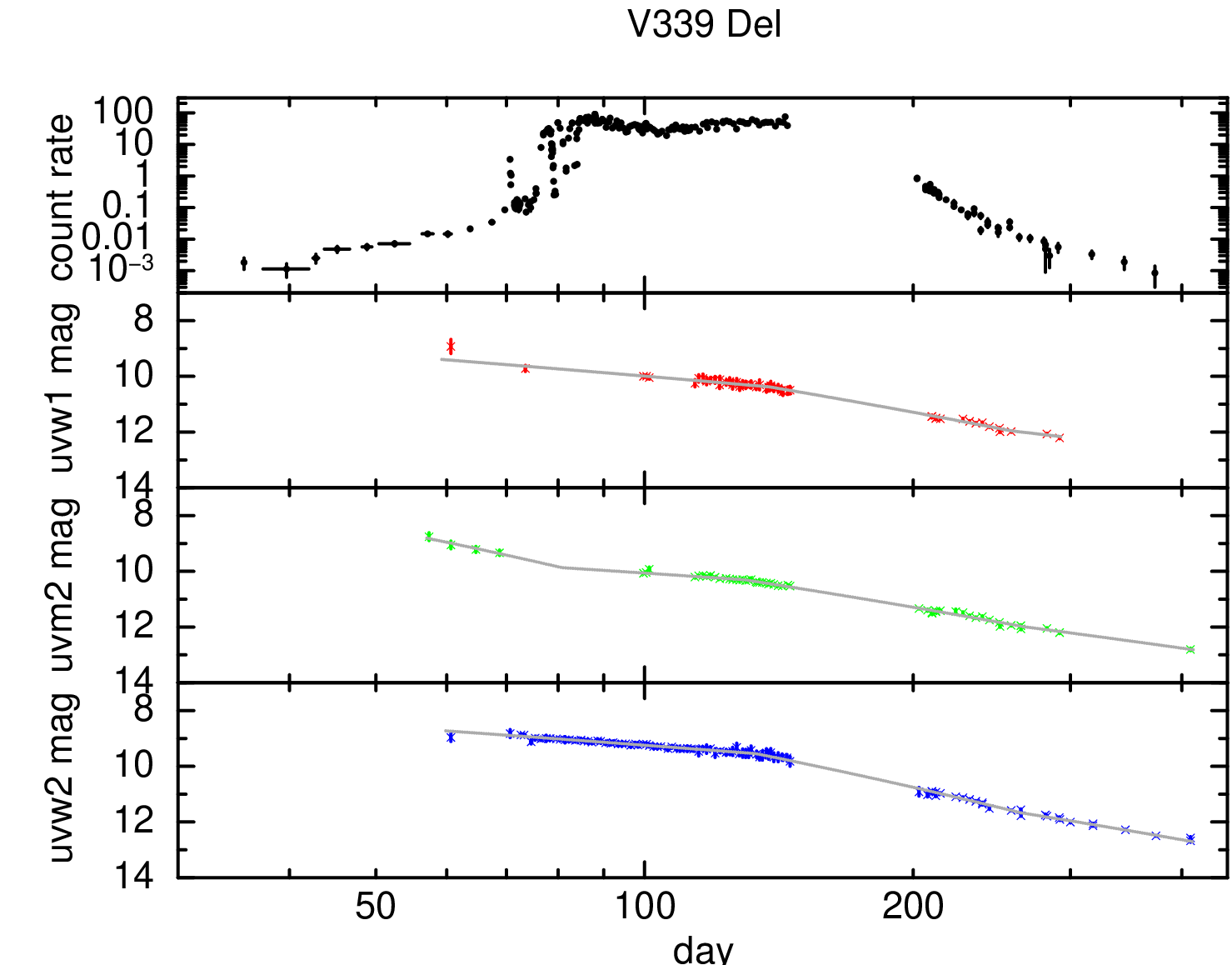}
\caption{The XRT and UVOT light-curves of V339 Del, with the best-fit broken power-law declines shown as the grey solid lines.  The X-ray data (top) are in count~s$^{-1}$ over 0.3--10~keV. \label{fig:V339Del}}
\end{figure}

\begin{figure}[H]
\includegraphics[width=10.5 cm,angle=-90]{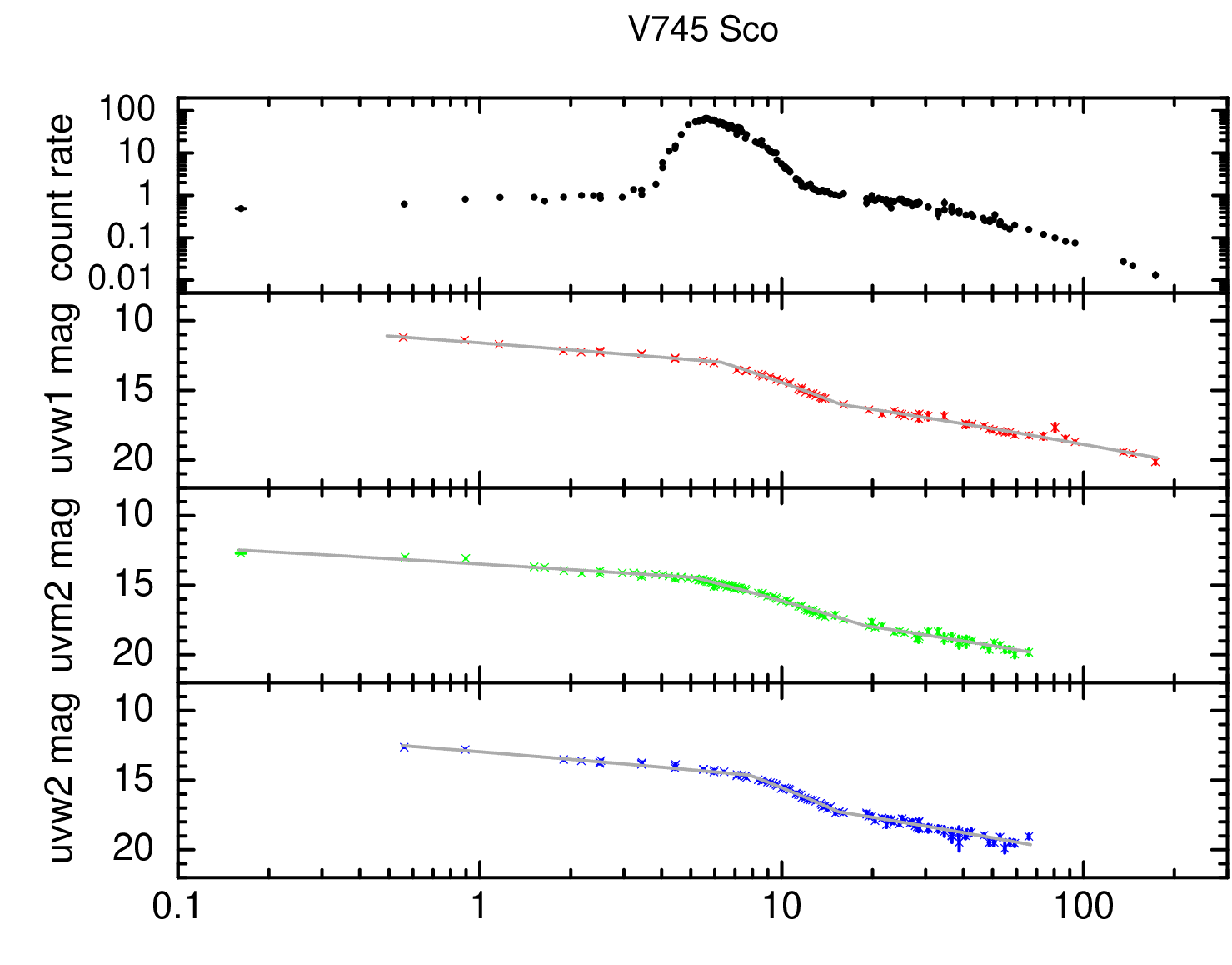}
\caption{The XRT and UVOT light-curves of V745 Sco, with the best-fit broken power-law declines shown as the grey solid lines. The X-ray data (top) are in count~s$^{-1}$ over 0.3--10~keV. \label{fig:V745Sco}}
\end{figure}

\begin{figure}[H]
\includegraphics[width=10.5 cm,angle=-90]{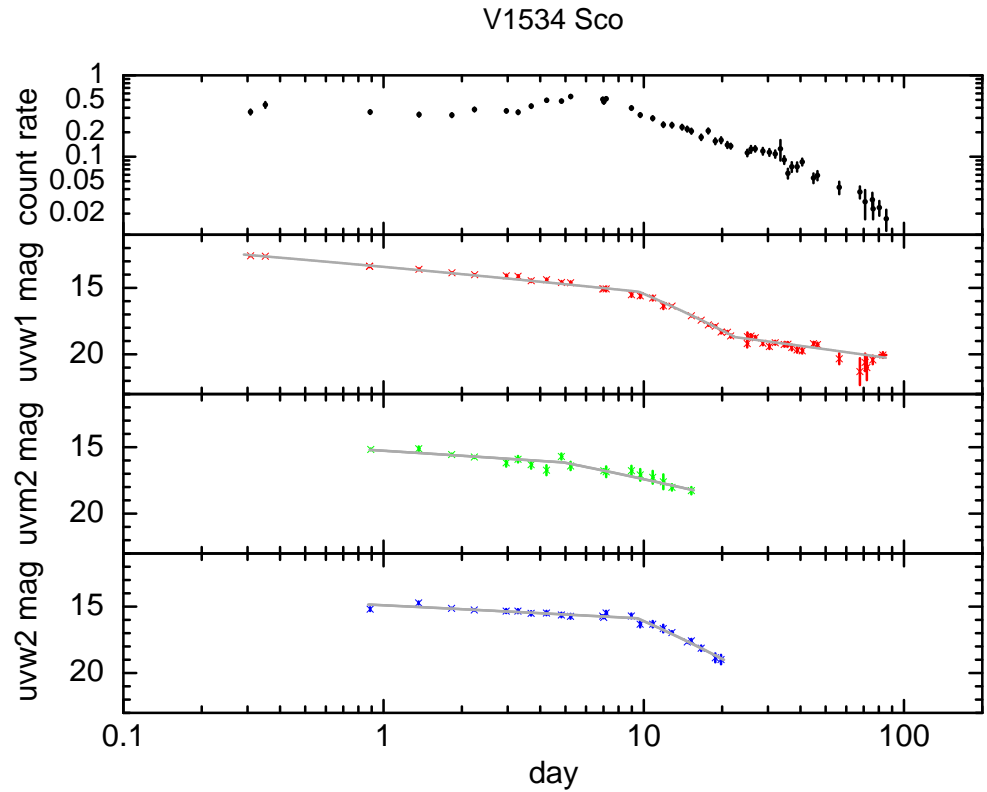}
\caption{The XRT and UVOT light-curves of V1534 Sco, with the best-fit broken power-law declines shown as the grey solid lines. The X-ray data (top) are in count~s$^{-1}$ over 0.3--10~keV. \label{fig:V1534Sco}}
\end{figure}

\begin{figure}[H]
\includegraphics[width=8.0 cm,angle=-90]{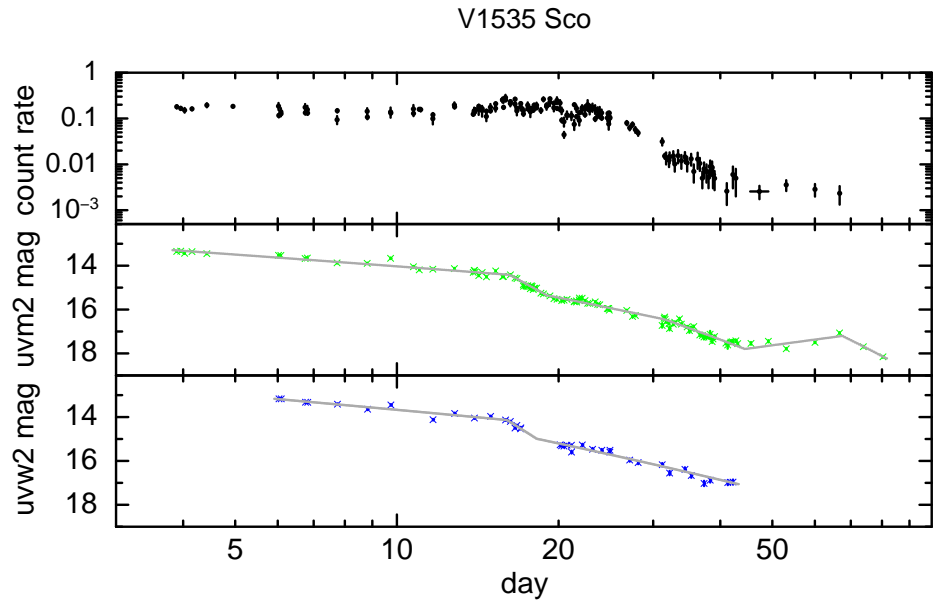}
\caption{The XRT and UVOT light-curves of V1535 Sco, with the best-fit broken power-law decline shown as the grey solid line. Data were only collected with the $uvw1$ filter intermittently for the first 19 days, so have not been included. The X-ray data (top) are in count~s$^{-1}$ over 0.3--10~keV. \label{fig:V1535Sco}}
\end{figure}

\begin{figure}[H]
\includegraphics[width=6.0 cm,angle=-90]{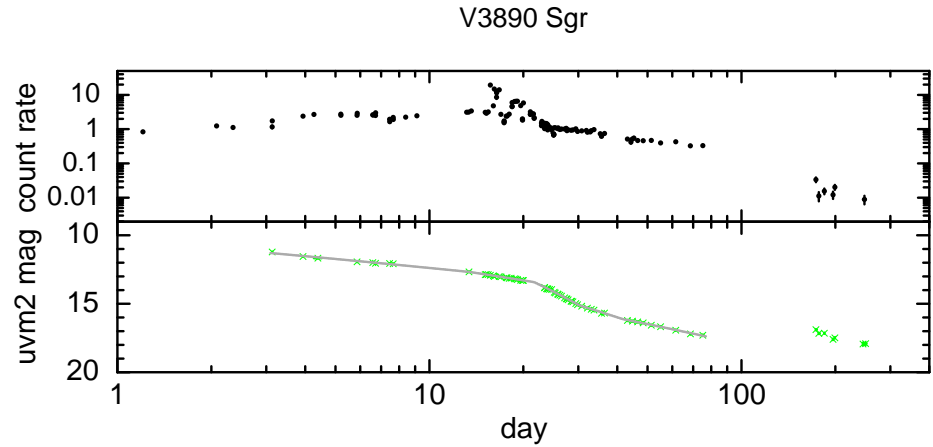}
\caption{The XRT and UVOT light-curves of V3890 Sgr, with the best-fit broken power-law decline shown as the grey solid line. $uvw2$ data were only collected until day 25, so have not been included. Late-time data after day 100 have been excluded from the fits. The X-ray data (top) are in count~s$^{-1}$ over 0.3--10~keV. \label{fig:V3890Sgr}}
\end{figure}

\begin{figure}[H]
\includegraphics[width=10.5 cm,angle=-90]{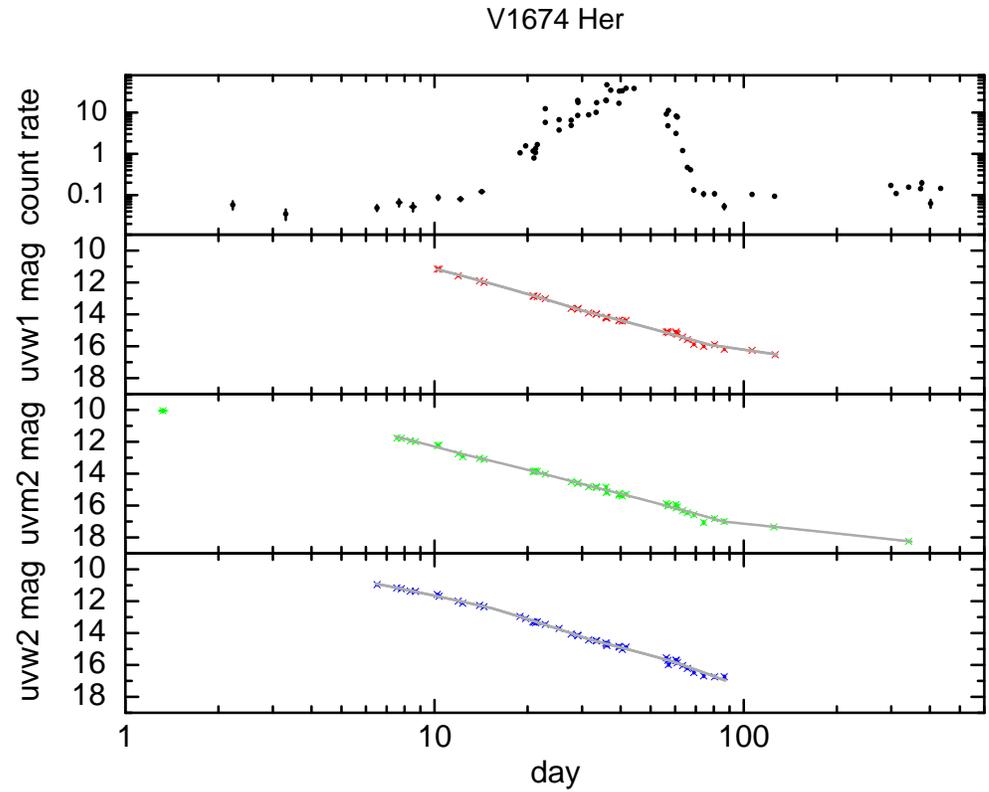}
\caption{The XRT and UVOT light-curves of V1674 Her, with the best-fit broken power-law declines shown as the grey solid lines. The earliest bin in the $uvm2$ light-curve has been excluded from the fit, since the break time cannot be constrained. The X-ray data (top) are in count~s$^{-1}$ over 0.3--10~keV. \label{fig:V1674Her}}
\end{figure}

\begin{figure}[H]
\includegraphics[width=8.5 cm,angle=-90]{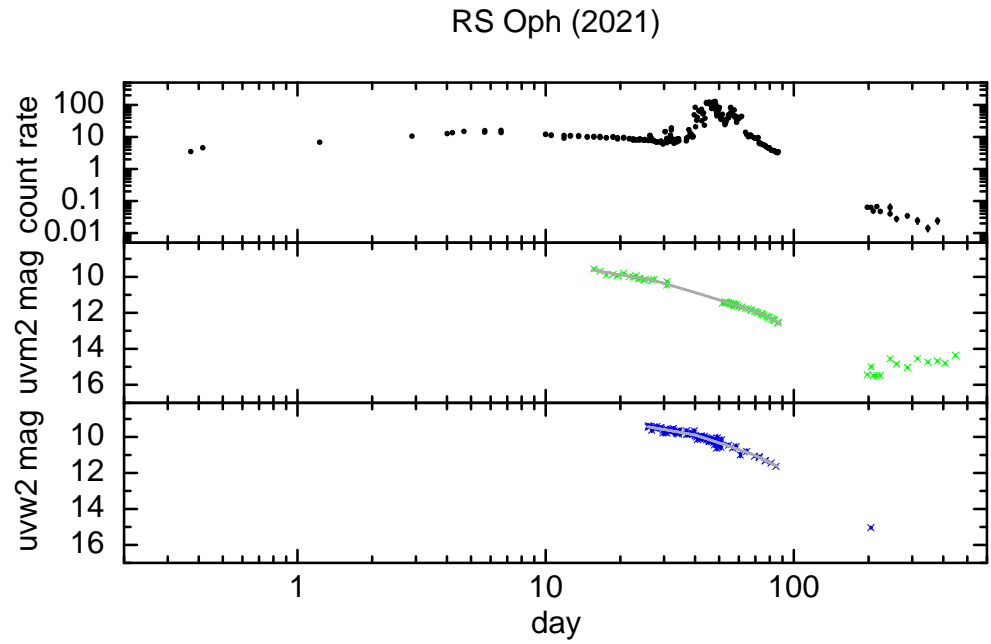}
\caption{The XRT and UVOT light-curves of RS Oph, with the best-fit broken power-law declines shown as the grey solid lines. Only the UVOT data before the first solar conjunction have been fitted. The X-ray data (top) are in count~s$^{-1}$ over 0.3--10~keV. \label{fig:RSOph}}
\end{figure}

Figs.~\ref{fig:V2491Cyg}--\ref{fig:RSOph} show the X-ray and UV light-curves for each of the novae in the sample, plotted as days since eruption.  The solid grey lines show the broken power-laws fitted to the UVOT data; the corresponding power-law indices for each segment are listed in Table~\ref{tab:fits}. Where the value of $\alpha$ is negative, this indicates that the source has rebrightened. The error bars on the data points are mostly smaller than the marker size, and are usually $\sim$~0.02--0.05 mag, occasionally up to $\sim$~0.3 mag for some of the shorter (and fainter) observations.

\begin{table}[H] 
\caption{Parameterising the UV light-curves as a series of power-law decays. The uncertainties on the fitted break times are given at the 90\% level, and the power-law slopes are calculated using Equation~\ref{eqn}. The errors on the slopes have been estimated using the uncertainties of the magnitudes from the fits. \label{tab:fits}}
\begin{adjustwidth}{-\extralength}{0cm}
\begin{tabularx}{535pt}{XCCXCXCXCXCXC}
\toprule
\textbf{Nova} & \textbf{filter}	& \textbf{$\alpha_1$} & \textbf{T$_{\rm break,1}$} & \textbf{$\alpha_2$} & {T$_{\rm break,2}$} & \textbf{$\alpha_3$} & \textbf{T$_{\rm break,3}$} & \textbf{$\alpha_4$} & \textbf{T$_{\rm break,4}$} & \textbf{$\alpha_5$} & {T$_{\rm break,5}$} & {$\alpha_6$}  \\
& & & (day) & & (day) & & (day) & & (day) & & (day)\\
\midrule
V2491  & $uvw2$ & 1.93$^{+0.02}_{-0.02}$ & 40.2$^{+0.1}_{-0.1}$ & 4.60$^{+0.01}_{-0.01}$ & 55.1$^{+0.1}_{-0.1}$ & 2.91$^{+0.09}_{-0.09}$ & 124$^{+2}_{-3}$ & 1.5$^{+0.1}_{-0.1}$ \\
Cyg\\
\midrule
U~Sco & $uvw1$ & 0.54$^{+0.05}_{-0.05}$ & 4.4$^{+0.1}_{-0.3}$ & 3.0$^{+0.1}_{-0.1}$& 12.52$^{+0.01}_{-0.01}$ & $-$0.28$^{+0.01}_{-0.01}$ & 24.19$^{+0.06}_{-0.06}$ & 4.99$^{+0.01}_{-0.01}$ & 44.51$^{+0.06}_{-0.14}$ & $-$2.85$^{+0.06}_{-0.06}$ & 52.28$^{+0.07}_{-0.03}$ & 8.34$^{+0.02}_{-0.02}$ \\
 & $uvw2$ & 0.44$^{+0.02}_{-0.02}$ & 3.9$^{+0.1}_{-0.1}$ & 3.20$^{+0.02}_{-0.02}$ & 12.53$^{+0.01}_{-0.09}$ & 0.02$^{+0.01}_{-0.01}$ & 25.15$^{+0.01}_{-0.06}$ & 5.61$^{+0.02}_{-0.02}$ & 41.8$^{+0.1}_{-0.2}$ & $-$0.33$^{+0.04}_{-0.04}$ & 52.34$^{+0.02}_{-0.07}$ & 7.56$^{+0.02}_{-0.02}$\\
\midrule
V407  & $uvm2$ & 0.35$^{+0.01}_{-0.01}$ & 41.4$^{+0.3}_{-0.4}$ & 4.39$^{+0.07}_{-0.07}$ & 63$^{+1}_{-1}$ & 2.23$^{+0.06}_{-0.06}$\\
Cyg\\
\midrule
T~Pyx & $uvm2$ & 6.54$^{+0.01}_{-0.01}$ & 113.90$^{+0.01}_{-0.01}$ & 1.27$^{+0.01}_{-0.01}$ & 142.59$^{+0.01}_{-0.01}$ & 7.03$^{+0.05}_{-0.05}$ & 150.2$^{+0.2}_{-0.3}$ & 4.05$^{+0.01}_{-0.01}$ & 268.7$^{+0.9}_{-0.8}$ & 1.38$^{+0.02}_{-0.02}$\\
\midrule
V959 & $uvw1$ & 2.02$^{+0.05}_{-0.05}$ & 212$^{+1}_{-1}$ & 5.33$^{+0.08}_{-0.08}$ \\
Mon  & $uvm2$ & 1.83$^{+0.01}_{-0.01}$  & 205$^{+1}_{-1}$ & 4.91$^{+0.07}_{-0.07}$\\
 & $uvw2$ & 2.07$^{+0.04}_{-0.04}$ & 205$^{+1}_{-1}$ & 5.29$^{+0.06}_{-0.06}$ \\
\midrule
V339 & $uvw1$ &  --& -- & 1.1$^{+0.2}_{-0.2}$ & 138$^{+16}_{-14}$ & 2.4$^{+0.2}_{-0.2}$ & 258$^{+12}_{-4}$ & 1.5$^{+0.5}_{-0.5}$\\
Del & $uvm2$ & 3.0$^{+0.5}_{-0.5}$ & 81$^{+15}_{-7}$ & 0.8$^{+0.5}_{-0.2}$ & 129$^{+2}_{-1}$ & 2.17$^{+0.07}_{-0.07}$ & 266$^{+11}_{-9}$ & 1.7$^{+0.2}_{-0.2}$\\
& $uvw2$ & -- & -- & 0.94$^{+0.08}_{-0.08}$ & 133$^{+5}_{-4}$ & 2.96$^{+0.02}_{-0.02}$ & 264$^{+2}_{-6}$ & 2.1$^{+0.1}_{-0.1}$\\
\midrule
 V745  & $uvw1$ & 0.68$^{+0.01}_{-0.01}$ & 6.3$^{+0.1}_{-0.1}$ & 3.01$^{+0.05}_{-0.05}$ & 15.7$^{+0.5}_{-0.4}$ & 1.45$^{+0.03}_{-0.03}$\\
 Sco  & $uvm2$ & 0.53$^{+0.01}_{-0.01}$ & 5.5$^{+0.1}_{-0.1}$ & 2.4$^{+0.1}_{-0.1}$ & 18$^{+1}_{-1}$ & 1.40$^{+0.03}_{-0.03}$\\
  & $uvw2$ & 0.73$^{+0.01}_{-0.01}$ & 7.8$^{+0.1}_{-0.1}$ & 3.62$^{+0.06}_{-0.06}$ & 15.5$^{+0.4}_{-0.4}$ & 1.46$^{+0.04}_{-0.04}$\\
\midrule
  V1534 & $uvw1$ & 0.75$^{+0.02}_{-0.02}$ & 9.5$^{+0.4}_{-0.4}$ & 3.65$^{+0.06}_{-0.06}$ & 22$^{+1}_{-1}$ & 1.1$^{+0.1}_{-0.1}$\\
  Sco & $uvm2$ & 0.5$^{+0.1}_{-0.1}$ & 4.9$^{+1.6}_{-0.6}$ & 1.7$^{+0.2}_{-0.2}$ & -- & --\\
  & $uvw2$ & 0.43$^{+0.09}_{-0.09}$ & 9.4$^{+1.1}_{-0.7}$ & 3.7$^{+0.3}_{-0.3}$ & -- & --\\
\midrule
  V1535 & $uvm2$ & 0.71$^{+0.01}_{-0.01}$ & 16.2$^{+0.1}_{-0.1}$ & 5.20$^{+0.01}_{-0.01}$ & 19.0$^{+0.2}_{-0.1}$ & 2.0$^{+0.1}_{-0.1}$ & 31.4$^{+0.3}_{-0.8}$ & 3.7 $^{+0.2}_{-0.2}$& 44.4$^{+0.6}_{-0.6}$ & $-$1.32$^{+0.02}_{-0.02}$ & 67$^{+1}_{-1}$ & 5.00$^{+0.06}_{-0.06}$\\
  Sco & $uvw2$ & 0.85$^{+0.05}_{-0.05}$ & 16.0$^{+0.2}_{-0.2}$ & 6.4$^{+0.4}_{-0.4}$ & 18.2$^{+0.6}_{-0.4}$ & 2.2$^{+0.1}_{-0.1}$ & -- & -- & -- & -- & -- & --\\
\midrule
  V3890 & $uvm2$ & 0.88$^{+0.06}_{-0.06}$ & 13$^{+1}_{-1}$ & 1.4$^{+0.1}_{-0.1}$ & 21.6$^{+0.2}_{-0.2}$ & 4.7$^{+0.1}_{-0.1}$ & 30.2$^{+0.6}_{-0.6}$ & 2.9$^{+0.3}_{-0.3}$ & 41$^{+3}_{-3}$& 2.1$^{+0.1}_{-0.1}$\\
 Sgr\\
\midrule
  V1674  & $uvw1$ & --  & -- & 2.2$^{+0.2}_{-0.2}$ & 30$^{+2}_{-4}$ & 2.1$^{+0.1}_{-0.1}$ & 77$^{+8}_{-3}$ & 1.2$^{+0.2}_{-0.2}$\\
Her  & $uvm2$ & 2.1$^{+0.2}_{-0.2}$ & 12.3$^{+1.2}_{-0.3}$ & 1.96$^{+0.01}_{-0.01}$ & 32$^{+8}_{-5}$ & 2.02$^{+0.01}_{-0.01}$ & 86$^{+6}_{-5}$ & 0.80$^{+0.06}_{-0.06}$\\
& $uvw2$ & 1.6$^{+0.2}_{-0.2}$ & 15$^{+1}_{-1}$ & 2.6$^{+0.1}_{-0.1}$ & 30$^{+2}_{-3}$ & 2.2 $^{+0.2}_{-0.2}$& -- & --\\
 \midrule
 RS & uvm2 & 0.99$^{+0.08}_{-0.08}$ & 27$^{+1}_{-1}$ & 1.70$^{+0.03}_{-0.03}$ & 76$^{+2}_{-3}$ & 3.6$^{+0.6}_{-0.6}$ \\
 Oph & uvw2 & 1.1$^{+0.2}_{-0.2}$ & 40$^{+3}_{-2}$ & 1.8$^{+0.6}_{-0.6}$ & 68$^{+17}_{-9}$ & 2.6$^{+0.10}_{-0.10}$\\
\bottomrule
\end{tabularx}
\end{adjustwidth}
\end{table}
\unskip

\subsection{Light-curve descriptions}

Below we briefly summarise the fits for each nova. We note that the models applied may not be a unique description of the data. For example, in some cases consecutive slope changes are in the same direction -- i.e., we see a gradual flattening across two break times, rather than a flattening followed by a steepening -- which may suggest a smoother, slower change is occurring, rather than the instantaneous breaks we fit. The series of sharply-broken power-laws is, however, a simple and convenient characterisation of the decline in UV brightness.

\subsubsection{V2491 Cyg}
Data were collected in the $uvw2$ filter between days 7 and 236; the earliest magnitudes were estimated using the read-out streak method \cite{readout}   (Fig.~\ref{fig:V2491Cyg}). The UV light-curve shows a rebrightening starting around day 12 (a `cusp'; \cite{strope10}), so only the data after day 19 were fitted. The data before the start of the cusp align with an extrapolation of the earliest fitted power-law, however. During the post-cusp time interval, the rate of the UV decline steepened around day 40, before starting to flatten off at day 55 and flattening further after day 133.

\subsubsection{U Sco}
\begin{figure}[H]
\includegraphics[width=6.0cm,angle=-90]{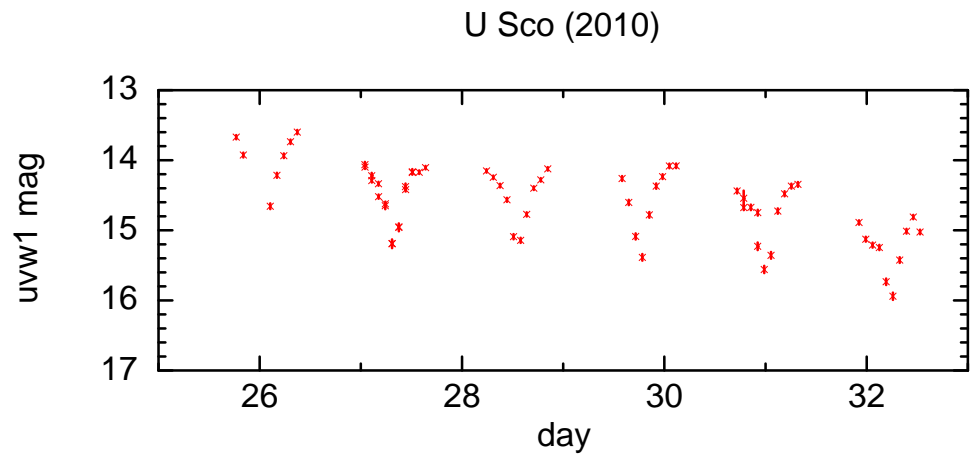}
\caption{Zoom-in of part of the U~Sco $uvw1$ light-curve, showing a series of eclipses. \label{fig:eclipses}}
\end{figure}

Following the 2010 eruption of the RN U~Sco, the majority of the UVOT data were collected using the $uvw1$ and $uvw2$ filters, between days 0.6 and 63 post-eruption; the earliest magnitudes were estimated using the read-out streak method \cite{readout} (Fig.~\ref{fig:USco2010}). U~Sco is an eclipsing system \citep[e.g., ][]{brad10c, ness12, ash15}, with the 1.23~d period clearly visible in the UVOT data (Fig.~\ref{fig:eclipses}). However, the underlying long-term evolution (including all the data within and outside the eclipse times) can be approximated by a series of power-law segments, showing both fading and flattening/brightening intervals. The fitted break times in the light-curves agree between the filters to within 
a few days. 

\subsubsection{V407 Cyg}

Between days 5 and 141, observations were performed using the $uvm2$ filter (Fig.~\ref{fig:V407Cyg}). During this time, the decay steepened on day 41, before flattening off again after day 63.

\subsubsection{T Pyx}

T~Pyx was observed while the UV emission was still rising; these early data, starting on day 7, have been excluded, with the decay fitting performed after day 70, once the emission had peaked, and running through till day 369. The $uvm2$ light-curve (Fig.~\ref{fig:TPyx}) showed a complex decay, with at least five separate segments, including a short, week-long interval between days 142 and 149, which showed a significant steepening.

\subsubsection{V959 Mon}

V959~Mon was initially monitored only using the $uvm2$ filter between days 58 and 150; after this time, the other two UV filters were also brought into play (Fig.~\ref{fig:V959Mon}). There was no notable change in decay rate at the earlier times, and all three filters show breaks from a flatter to steeper decay around day 205-212.

Only data until day 256 are considered; after this time there were observations on days 423 (all three filters), 687 ($uvw1$) and 1002 ($uvm2$), which do not provide enough data points to constrain additional fitting. The measurements do suggest a further flattening of the decay between days 256 and 423, however.

\subsubsection{V339 Del}

The very earliest observations of V339~Del with {\em Swift} were performed with  the UVOT blocked, because of the extreme brightness (it peaked at $V$~$\sim$~4.3; \cite{munari15}); the UV data collection commenced on day 57 in the $uvm2$ filter, and day 60 in $uvw1$ and $uvw2$, continuing until day 291 in $uvw1$, and day 409 in both $uvm2$ and $uvw2$ (Fig.~\ref{fig:V339Del}). Again, the read-out streak method \cite{readout} was required for data brighter than around magnitude 10--11. Since the observations were performed mainly with the high time resolution event mode, data were subsequently averaged over bins of $\sim$~1~d in length, to improve the statistics. The $uvm2$ filter was the one used most frequently at the start, and this higher density of early data points is likely why an additional early-time break, from steep to flat can be constrained in this light-curve. All three UV light-curves subsequently steepen around day 130--140, before flattening between day 258--266.

\subsubsection{V745 Sco}

V745~Sco was observed with all three filters from 0.5 day after eruption until day 66. There was also an earlier $uvm2$ observation at day 0.16, and further observations out till day 173 in $uvw1$ (Fig.~\ref{fig:V745Sco}). All three light-curves steepen around day 5--8, and flatten off again between $\sim$~day~16--19. 

\subsubsection{V1534 Sco}

While V1534~Sco was initially observed with all UV filters, beyond three weeks after eruption only the $uvw1$ observations were continued (Fig.~\ref{fig:V1534Sco}). Considering the longest ($uvw1$) dataset, which runs between days 0.3 and 83, the decay broke twice, first steepening after day 9.5, then flattening again after day 22. The $uvm2$ and $uvw2$ light-curves only cover the time of the first break, with data collected between days 0.9--15.2 and 0.9--19.8, respectively. While the break time fitted to the $uvw2$ data, and the post-break slope, are in close agreement with $uvw1$, the initial slope is flatter. The break in $uvm2$ appears to occur days earlier.

\subsubsection{V1535 Sco}

V1535~Sco was observed between days 3.8 and 80.3 with the $uvm2$ filter, and 6.0 to 42.2 with $uvw2$ (Fig.~\ref{fig:V1535Sco}). The $uvw1$ filter was only used for seven observations, so these data have not been included. Because of the longer duration of the $uvm2$ dataset, it is better fitted with a further three breaks, in addition to the two which are required for the $uvw2$ light-curve, around 16 and 18--19~days. After day 44, the $uvm2$ data show a brightening trend for around 23 days, before starting to fade again.

\subsubsection{V3890 Sgr}

The vast majority of the UVOT observations of V3890~Sgr were obtained using the $uvm2$ filter, so these are the data considered here (Fig.~\ref{fig:V3890Sgr}). Observations in this filter began on day 3, continuing until the nova entered the solar observing constraint for {\em Swift} on day 75. Further observations were taken after the source re-emerged from behind the Sun, until day 250, but these have not been included in the overall fit (though are shown in Fig.~\ref{fig:V3890Sgr}). The light-curve decline steepens twice and then flattens off again, in two stages. The UV brightness appears to have stayed constant during the observing constraint (or the source faded and rebrightened by about the same amount), before starting to fade again, with $\alpha$~$\sim$~2.8. We note that this flat interval is much later than the X-ray SSS phase, so will not correspond to the plateau phase of \cite{hachisu08}.

\subsubsection{V1674 Her}

An early observation was obtained on day 1.3 using the $uvm2$ filter, with following observations on days 6.5, 7.6 and 10.3 in $uvw2$, $uvm2$ and $uvw1$ respectively, before subsequent observations used all three filters until day 86, after which filters were again alternated (Fig.~\ref{fig:V1674Her}). The last observations in $uvw1$, $uvm2$ and $uvw2$ were on days 126.5, 341.4 and 86.5, respectively. 
There is clearly a change in slope between the first two observations obtained with the $uvm2$ filter on days 1.3 and 7.6, but a break time cannot be sensibly estimated with no measurements in between. Therefore the initial $uvm2$ data point was not included in the analysis. 

While the $uvm2$ data show evidence for three breaks in the power-law decay, the $uvw1$ data do not require the earliest break, most likely since this filter dataset starts latest. Similarly, the $uvw2$ light-curve does not require the final change in slope because of the lack of data beyond day 86. Although $\alpha_2$ and $\alpha_3$ are very similar for the $uvm2$ data, the slight change is an improvement (the break is significant at the 3$\sigma$ level), and, given the requirement for a break around day 30 in the other two filters, it seems likely to be real.

\subsubsection{RS Oph}

The very earliest {\em Swift} observations of RS~Oph in 2021 had the UVOT blocked, because of the optical brightness of the source. The first photometric observation was obtained 15.6 days after the optical peak, and ran through till the solar observing constraint began on day 87; observations began again on day 197.5 and continued until the following solar constraint, on day 448. The read-out streak method \cite{readout} was used for the first two months. The majority of observations were obtained with either the $uvm2$ or $uvw2$ filter (sometimes with grism exposures as well; \cite{azz22}), and these are shown in Fig.~\ref{fig:RSOph}.  Both filters show two steepening breaks, the first around day 30--40, the second, day 70--75. Following solar conjunction, the $uvm2$ light-curve shows signs of slow rebrightening; these data have not been included in the fits.

\subsection{Normalised break times}

In order to compare the different novae more directly, in Table~\ref{tab:norm} the break times have been normalised to the end of the X-ray SSS phase as T$_{\rm break}$/T$_{\rm SSSend}$. Each SSS end point has been estimated from the X-ray light-curves (Figs.~\ref{fig:V2491Cyg}-\ref{fig:RSOph}; see also \cite{page20a, page20b, drake21}), and is defined to be the time at which the X-ray count rate starts to decrease steadily. There can obviously be some uncertainty in this time, given that there are gaps in the light-curve coverage, as well as short-term variability, but this is typically only a couple of days or less\footnote{The SSS phase of V339~Del ended during the solar observing constraint, when no observations could be performed. However, an extrapolation backwards of the decline once observations restarted suggests the X-ray fading began around the same time as the constraint.}, as can be seen from the top panels of Figs.~\ref{fig:V2491Cyg}--\ref{fig:RSOph}.

\begin{table}[H] 
\caption{The nth break time normalised to the end time of the X-ray SSS: T$_{\rm break,n}$~(norm.)~=~T$_{\rm break,n}$/T$_{\rm SSSend}$. The power-law indices are the same as in Table~\ref{tab:fits}, so have not been listed. \\$^{a}$ The SSS switched off during the Sun constraint, which ran from days 144--202; given the shape of the decay, it seems likely the SSS ended around the time V339~Del entered the observing constraint.\\
$^b$ In V1534~Sco, the early brightening of the X-ray emission appears to be caused by the absorbing column declining, rather than a new soft component appearing. The X-ray count rate peaked on day 7, and this is the value used for the break time normalisation. \label{tab:norm}}
\begin{tabularx}{400pt}{XXCCCCCC}
\toprule
\textbf{Nova} &  \textbf{T$_{\rm SSSstart}$--T$_{\rm SSSend}$} &   \textbf{filter}	&  \textbf{T$_{\rm break,1}$}  & \textbf{T$_{\rm break,2}$} & \textbf{T$_{\rm break,3}$}  & \textbf{T$_{\rm break,4}$} &  \textbf{T$_{\rm break,5}$}   \\
&\textbf{(day)} &  &  \textbf{(norm.)}  & \textbf{(norm.)}  & \textbf{(norm.)} &  \textbf{(norm.)}  & \textbf{(norm.)}   \\

\midrule
V2491  & 33--43  &  $uvw2$ &  0.9  & 1.3 &  2.9  \\
Cyg\\
\midrule
U~Sco & 12--33  &  $uvw1$ & 0.1 &  0.4 &  0.7 &  1.4 &  1.6  \\
 & &   $uvw2$ &  0.1  & 0.4 &  0.8 &  1.3 &  1.6  \\
\midrule
V407  & 12--40 &   $uvm2$ &  1.0 &  1.6 & \\
Cyg\\
\midrule
T~Pyx & 123--180 &  $uvm2$ &  0.6  & 0.79  & 0.83 &  1.5 \\
\midrule
V959 & 150--200 &  $uvw1$ &  1.1 \\
Mon  & &  $uvm2$ &  1.0\\
 & &  $uvw2$ &  1.0 \\
\midrule
V339 & 60--144$^{a}$ & $uvw1$ & -- &  0.9 &  1.8 \\
Del &  & $uvm2$ &  0.6  & 0.9 &  1.8 \\
& & $uvw2$ & --&   0.9 &  1.8 \\
\midrule
 V745  & 3--6 &  $uvw1$ &  1.1  & 2.6 \\
 Sco  & &  $uvm2$ &  0.9 &  3.2 \\
  & &  $uvw2$ &  1.3  & 2.6 \\
\midrule
 V1534 & 7$^b$ &  $uvw1$ &  1.4 &  3.2 \\
  Sco & & $uvm2$  &  0.7 & --\\
& & $uvw2$ &  1.2 & -- \\
\midrule
 V1535 & 12--25 &  $uvm2$ &  0.6 &  0.8 &  1.3 &  1.8 &  2.7 \\
  Sco & &  $uvw2$ &  0.6 &  0.7 & -- & -- & --\\
\midrule

 V3890 & 8--20 &  $uvm2$ &  0.7 &  1.1 &  1.5 &  2.1 \\
 Sgr\\
\midrule
  V1674  & 19--45 &  $uvw1$ & -- &  0.7 &  1.7 \\
Her  & &  $uvm2$ &  0.3 &  0.7 &  1.9 \\
& &  $uvw2$ &  0.3 &  0.7 & --\\
\midrule
RS & 21--62 & uvm2 &  0.4 &  1.2  \\
 Oph &  & uvw2 &  0.6  & 1.1\\
 
\bottomrule
\end{tabularx}
\end{table}
\unskip

\begin{figure}[H]
\includegraphics[width=9 cm,angle=-90]{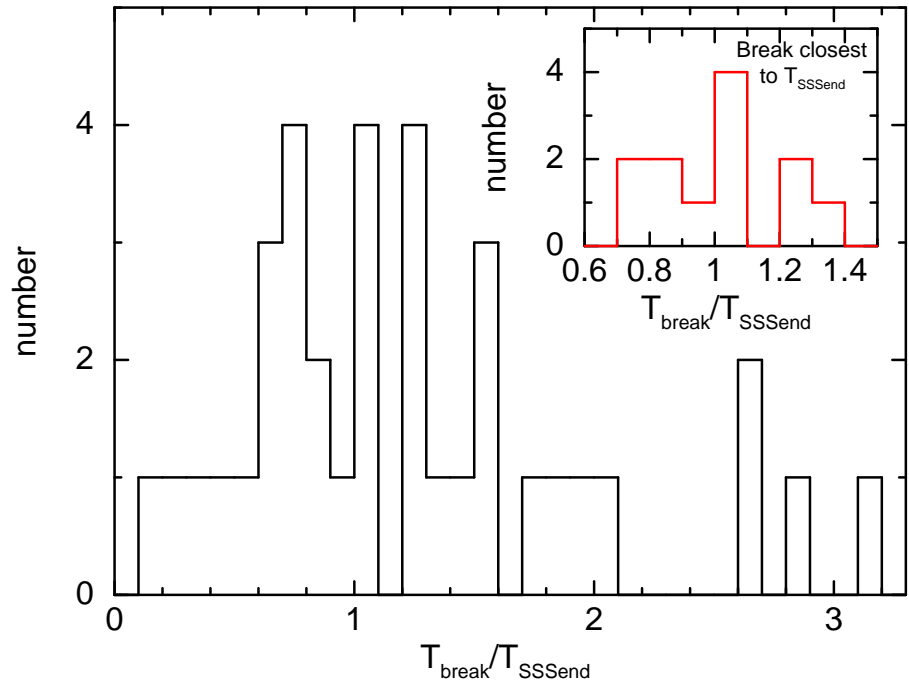}
\caption{A histogram showing the normalised break times fitted in the UVOT light-curves for all novae in the sample; the best-sampled light-curve has been used in each case where observations were performed in more than one filter. The inset plots only the break for each nova light-curve which is closest in time to T$_{\rm break}$/T$_{\rm SSSend}$~=~1. \label{fig:histo}}
\end{figure}

Each light-curve shows a break close in time to the end of the SSS phase (i.e., there is a T$_{\rm break}$ (norm) close to 1 -- in the range 0.7--1.3 -- in Table~\ref{tab:norm}), with the possible exception of V1534~Sco, which does not break until a little later in the $uvw1$ band. 
However, as mentioned in the caption of Table~\ref{tab:norm}, V1534~Sco did not show an obvious SSS phase, but rather became softer and brighter as the absorbing column decreased, so this discrepancy (and the mis-match between the break times in the different filters) can be disregarded. 
Fig.~\ref{fig:histo} shows the normalised break times (using the best-sampled light-curve for each nova which was observed using more than one UVOT filter) plotted as a histogram. The inset shows only the break which is closest in time to the end of the SSS phase, since some of the light curves have multiple breaks around this interval. There is a clear peak in the number of breaks around T$_{\rm SSSend}$. The fact that some breaks occur before the apparent end of the SSS emission in the X-ray band will be returned to in Section~\ref{sec:comp}.

\section{Discussion}

The UV light-curves presented here were chosen to be well sampled, and it is obvious that a higher cadence of observations is more likely to highlight changes in slope, especially over shorter time intervals\footnote{U Sco was observed at a particularly high cadence when trying to map the 1.23-day eclipses.}. This should be borne in mind when comparing the fits to different novae.

\subsection{Individual novae}

Two thirds of the novae in this sample have multi-filter light-curves. While there are distinct similarities between the different filters, the best fits are not always identical. 

Considering the break times normalised with respect to the end of the SSS in Table~\ref{tab:norm}, there is a slight difference in the third and fourth break times for U~Sco (with T$_{\rm break,4}$ being around 2.7~d earlier in $uvw2$ from Table~\ref{tab:fits}), causing different slopes in the plateau, fading and late re-brightening. 

V959~Mon shows a later break (by about a week) in $uvw1$, although the $uvw1$ and $uvw2$ slopes are very similar; $uvm2$ shows flatter declines both pre- and post-break.

In V745~Sco, the first break happens in the order $uvm2$, $uvw1$, $uvw2$; however, the second break time is consistent for $uvw1$ and $uvw2$, with $uvm2$ apparently breaking a couple of days later.
We note that the $uvw1$ and $uvw2$ filter transmission curves have extended red tails -- the so-called `red leak'\footnote{\url{https://swift.gsfc.nasa.gov/analysis/uvot_digest/redleak.html}}. This might in general affect the break times in these two filters compared with $uvm2$ (which does not suffer from this red leak) in systems with a red companion.

In the case of V1534~Sco, only the first break was covered by all three filters, and the break in $uvm2$ is noticeable earlier (around 4.5~d), leading to a flatter post-break decay slope. The longest dataset, $uvw1$, fades more steeply than the other two before this early-time break.

V1535~Sco has very similar break times for both filters used for the early monitoring. The slope between the breaks is steeper in the $uvw2$ data, however.

V1674~Her shows an earlier first break in $uvm2$ than $uvw2$ (no break required in $uvw1$ due to lack of early data). Around this break, the $uvm2$ light-curve flattens, whereas the $uvw2$ data steepen. The break around day 30 is consistent in all three filters, and the third break agrees between the two filters used at that time, though the error bars on the time are relatively large.

For RS~Oph, the first break occurs slightly earlier in $uvm2$ than $uvw2$, but the second break occurs a little later -- although the second break time is consistent within the uncertainties. The initial decay slopes are similar for both filters, though the final decay is steeper in $uvm2$.

In some of these situations, the differences in observation times (and sometimes cadences) between the UVOT filters may account for some of the discrepancies found. There do seem to be some intrinsic differences between the results at different wavelengths, but there are no obvious trends which are the same in each case: we don't always see that the bluer $uvw2$ data fade in a different manner from the redder $uvw1$, for example. As noted above, $uvw1$ and $uvw2$ are affected by a `red leak', which could be a more significant issue where the secondary star is redder. UV spectra of novae often show strong emission lines, which can vary over time. Such emission will affect the overall magnitudes in different filters by differing amounts, depending on the wavelengths of the lines, which may lead to changes in the apparent break times between filters.
We recall that \cite{page15} extended the fitting to the optical and infrared (IR) bands for V745~Sco, finding that the optical curves changed more gradually than the UV, while the IR was more strongly affected by light from the red giant (RG) secondary.

\subsection{Comparing the sample as a whole}
\label{sec:comp}

If the pre-day-70 data for T~Pyx, and the single early $uvm2$ bin for V1674~Her, are included, then, with the exception of the $uvm2$ data for V339~Del, all the light-curves in the sample show a first break from a flatter to a steeper decay. This initial break occurs before or at the time of the SSS switch-off in the X-ray band\footnote{V1534~Sco is a possible exception, as previously noted.}. 

Whether it is the earliest fitted break or not, all the novae (with obvious supersoft emission) show a change in slope around the end of the SSS phase (given the uncertainties in determining the exact SSS end time), as has been previously noted by, e.g., \cite{page10, page13, page15}; see also Fig.~\ref{fig:histo}. 
The end of the SSS phase corresponds to the point at which nuclear burning ceases, thus leading to a drop in the ionisation of the RG wind in the symbiotic-like systems, and a general decline in any UV emission coming directly from the hot WD photosphere. 
All the breaks are from flatter to steeper at this time, with the exception of V1674~Her, for which $uvw1$ and $uvw2$ both show a flattening; the $uvm2$ curve shows a (slight) steepening, however. T~Pyx shows two breaks in quick succession (only a week apart) close to 0.8~T$_{\rm SSSend}$, with the first showing a flat to steep transition.  The V1535~Sco $uvm2$ light-curve has two breaks relatively close in time to the SSS end point (0.76$\times$ and 1.26$\times$ T$_{\rm SSSend}$), with the second of these times corresponding to a steepening, which may suggest that is the break related to the end of the SSS phase (and is therefore the normalised break time included in the inset of Fig.~\ref{fig:histo}).  

Fig.~\ref{fig:histo} shows that, for some novae, the break closest in time to the end of the SSS phase actually occurs a little before T$_{\rm SSSend}$ measured from the X-ray light-curves. While it is not immediately obvious why the UV emission would start to fade before the nuclear burning switches off, it could be that fitting the light-curves with sharp breaks affects the results, given that some breaks, at least, might in reality be a smoother transition between power-law slopes. In addition, changes in the continuum spectral shape, or the emergence (or variation) of emission lines could affect the rate at which the overall UV brightness changes.

All the novae then show further breaks in their UV decline after the end of the SSS phase. There are limited later-time measurements for V959~Mon, but it is clear that the decay flattened again; there just isn't enough information to constrain the fit. Following the 100+ day gap caused by the solar conjunction (which started around 10--15 days after the SSS phase started to fade), the data for RS~Oph show a brightening trend. These later changes in slope do not occur at fixed times with respect to the SSS turn-off, neither do different novae show similar numbers of breaks over given time intervals. The earliest fitted break occurs at only 10\% of the SSS end time in U~Sco, corresponding to around day 4 post-eruption (followed by a further early break at 40\% of the SSS time -- day 12.5 -- corresponding to the start of the plateau interval, which then persists for around 12~days). The UV decay in V745~Sco also breaks early on, around day 6, though that already equates to the end of the SSS phase in this rapidly-evolving nova. We note that T~Pyx shows a short `wiggle' at the end of its month-long plateau (days 114--143), fading steeply with $\alpha$~$\sim$~7 for about a week, before continuing to decay with $\alpha$~$\sim$~4 for more than 100 days. 

The different novae were followed for differing lengths of time overall, with respect to the end of the SSS phase (or, indeed, in general: usually out till at least 100 days post-eruption, though). The duration of the monitoring was generally set by the brightness of the fading nova (either X-ray or UV), though sometimes will also have been affected by {\em Swift} observing constraints. This means that we cannot sensibly compare the numbers of late-time (i.e., after the end of the SSS emission) breaks.

There is a wide range of decay slopes seen, from the flat plateau phase in some RNe, to steep declines of $\alpha$~$>$~5. Two of the novae (U~Sco and V1535~Sco) show temporary rebrightenings at later times; in fact, both of these happen around day 40 post-eruption, though these equate to different normalised times with respect to the SSS end point. The UV emission of RS~Oph also rebrightens at late times; this rebrightening has not yet ended with the observations collected to date.

Besides the break in decay caused by the end of nuclear burning, the other observed changes in slope are likely to be related to physical processes in the expanding ejecta, such as declining density, or (before the end of the SSS phase) decreasing incident flux due to greater distance from the WD. In some novae, the accretion process may remain unsettled for quite some time after the eruption, rather than promptly relaxing back into a typical accretion disc \cite{mason21}, and this could affect the rate at which the UV emission fades.

\subsection{Populations}

\begin{figure}[H]
\includegraphics[width=9.5cm,angle=-90]{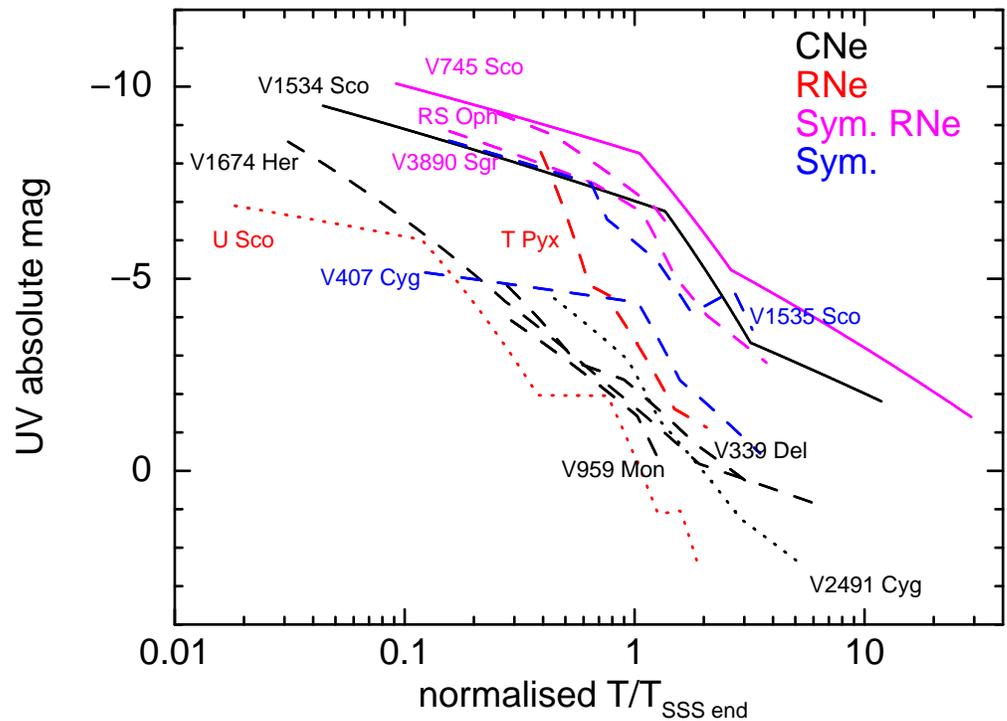}
\caption{Power-law fits to the UVOT light-curves, converted to absolute magnitude, and plotted against time normalised to the end of the SSS. In the case where a nova was observed in multiple filters, the best sampled curve is shown. Classical novae are plotted in black, short-period recurrent nova in red, long-period recurrent novae with RG companions in magenta, and non-recurrent symbiotic-like systems (those with giant companions) in blue. Solid lines mark the light-curves collected using the $uvw1$ filter, dashed -- $uvm2$ and dotted -- $uvw2$. \label{fig:all}}
\end{figure}

In Fig.~\ref{fig:all}, we show the power-law fits to the light-curves in this sample (one per nova, choosing the filter which covers the longest interval of time if more than one was used), transformed into absolute magnitude using the distances in Table~\ref{tab:novae}, and plotted against the time normalised to the SSS end. Corrections for reddening have also been applied at the wavelength of the UV filter in use each time, based on values of E(B$-$V) taken from the literature (mainly from \cite{brad22a}; see Table~\ref{tab:novae}), and assuming that A$_{\rm V}$ = 3.1~$\times$~E(B$-$V) for the Milky Way \cite{draine03}.
The early times for T~Pyx and V2491~Cyg which were excluded from the power-law fitting are not shown, likewise the V959~Mon and RS~Oph late time data.
The sub-samples of CNe, RNe with short periods, RNe with long periods, and symbiotic-like systems (which have not been seen to recur) are plotted in black, red, magenta and blue, respectively. Different line styles highlight which filter is plotted. The symbiotic RNe with long periods are at the brighter absolute magnitude end of the sample, with the CNe mainly at a much lower luminosity (V1534~Sco being the exception, though there is a possibility that this source is actually a symbiotic system); there do not appear to be any other obvious differences between the populations.

\subsubsection{Classical novae}

The classical (non-symbiotic) novae in this current sample are V2491~Cyg, V959~Mon, V339 Del, V1674~Her and (probably) V1534~Sco. The UV light-curves of V959~Mon started later after eruption because of the proximity to the Sun delayed observations, so early-time breaks in the decay may have been missed. Despite this, these four objects show similar power-law decays and breaks. 
The spread of unreddened, absolute magnitudes is not large, either.

\subsubsection{Recurrent novae}

The recurrent novae in this paper consist of both short-period (U~Sco and T~Pyx), and long-period (with a giant companion) systems (V745~Sco, V3890~Sgr and RS~Oph). In this sample of 12 novae, the long-period, symbiotic RNe are typically the overall brightest in terms of absolute magnitude (although we did not measure the peak for CN V339~Del, since the UVOT was operated in blocked mode for instrument safety). These measurements do, of course, rely on having accurate distance and extinction measurements. For example, if RS~Oph were at 1.6~kpc (the previously favoured distance) instead of 2.7~kpc, it would be about one magnitude fainter in absolute terms.

U~Sco and T~Pyx, the two short-period systems, both show early time breaks and the (oft-seen) plateau corresponding to the X-ray SSS interval. None of the long period, symbiotic systems in this sample shows a plateau in their UV decline. The earliest break in the V745~Sco light-curve corresponds to the end of the SSS phase, which occurred very early, running only between days 3--6 post-eruption; there might not have been time for an optical/UV plateau to form during this brief interval. V3890~Sgr showed a longer SSS duration of $\sim$~12~d, though this is still shorter than in U~Sco ($\sim$~21~d) and T~Pyx ($\sim$~57~d) so, again, this might account for the lack of plateau. While RS~Oph does not show an obvious plateau in these UV data, its optical light-curves do, as shown by \cite{strope10, greg11}.

\subsubsection{Symbiotic-like novae}

In this sample there are five systems which have cool RG companions: V407~Cyg, V1535~Sco, V745~Sco, V3890~Sgr and RS~Oph. The first two are CNe, having only been seen in eruption once, while the other three are known RNe. V1535~Sco has an orbital period of 50~d, while the other four have even longer P$_{\rm orb}$~$>$~1~yr. Following a nova explosion in these embedded systems, there will be shock interactions between the expanding ejecta and the RG wind. As well as producing hard X-rays \cite[e.g.,][]{bode06,nelson15}, these shocks also lead to UV line emission \cite{azz21, azz22}.

The presence or otherwise of a giant companion and accompanying wind does not directly have an effect on the end time of the X-ray SSS emission. In a summary paper of {\em Swift}-XRT observations of SSS novae, \cite{greg11} also showed that there is no correlation between orbital period (typically longer for the symbiotic-like novae) and the SSS turn-off time. The SSS interval is generally accepted to be related to the WD mass, with an early onset and short duration being signs of a massive WD \cite[e.g.,][]{kato97,yaron05,henze11,greg11} -- as expected for RNe. We do indeed see that the RNe U~Sco, V745~Sco and V3890~Sgr show early and short SSS phases; T~Pyx does not, but this system is known to be unusual, evolving more slowly than most other recurrents \cite{brad10}. The start of the SSS phase in RS~Oph is not quite as early as some, but has a well-defined beginning and end.

For the symbiotic-like sample here, V745~Sco has a very short SSS phase (ending on day 6), while the other four switched off after 20--60~d. The SSS phases of the non-symbiotic systems end between days 33--200\footnote{As noted in Table~\ref{tab:norm}, V1534~Sco does not show an obvious SSS phase, with day 7 actually corresponding to the time when the X-ray emission peaked due to the decrease of the absorption. This nova has therefore be omitted from this range.}. 

In the case of the symbiotic-like novae, there will be a RG wind `bubble' surrounding the system. The outer extent of this region depends on the recurrence time of the nova: the longer the time between eruptions (or if only one nova event has so far occurred), the further the RG wind can flow before disruption. Once the nova ejecta reach this limit (where the wind density becomes too low to be significant), the shock `breaks out' of the medium, which will likely lead to a drop in the UV flux as the shock-related emission decreases, and so a steepening in the light-curve decay. As an example, in the 2006 eruption of RS~Oph (not included in this sample, since the majority of the UVOT observations were taken using the grism; \cite{azz21, azz22}), the shock breakout time was found to occur around day 80 post-outburst \cite{anupama08}.

There is no obvious difference between the UV light-curve evolution for the symbiotic-like systems and the others in this sample (Fig.~\ref{fig:all}) -- although it is likely that one of the later breaks (possibly later than {\em Swift} has followed, and therefore not included in the plots above) in the UV decline in the symbiotic/embedded systems is related to the drop-off of the RG wind.

\subsection{Comparison with previous work}
\label{other}

Strope et al. \citep{strope10} published almost 100 detailed nova optical light-curves, including some of the sources in our sample, categorising the results depending on the shapes of the light-curves. Those which follow a series of power-law declines in magnitude-log(time) space, as in the sample considered here, are classified as `S' for smooth (or stereotypical), and constitute almost 40\% of their sample. As mentioned above, RNe often show a plateau in their optical/UV decay, coincident with the SSS phase seen in X-rays; these are included in the `P'-class by \citep{strope10}, meaning an `S'-type curve, but with a long, almost-flat interval a few magnitudes below peak. Considering the RNe in this sample, U~Sco and T~Pyx are indeed both marked as `P' by \cite{strope10}, and a flat interval is also seen in our fits (see Table~\ref{tab:fits}, and Figs.~\ref{fig:USco2010} and \ref{fig:TPyx}). RS~Oph is also classified as as `P', and does indeed show a plateau in the optical light-curve \cite{strope10}, but no such flattening is obvious in the UV data (Fig.~\ref{fig:RSOph}).
RN V3890~Sgr does not show a plateau, and is marked as `S' in \citep{strope10}; see also \cite{page20a}. RN V745~Sco is not classified by \citep{strope10}, but does not show a plateau in the UVOT data presented here. 

V2491~Cyg is one of only a few novae which have been found to show a cusp-shaped secondary maximum (`C' in the Strope classification scheme: ``power-law decline plus secondary maximum with steepening rise then steep decline''), which is clearly visible in the UVOT light-curve in Fig.~\ref{fig:V2491Cyg}.

We do note that the specific RN eruptions considered by \citep{strope10} are not the exact same events as those presented here, with our data corresponding to more recent outbursts. Optical/UV light-curves from recurrent eruptions are frequently very similar -- possibly identical -- for both Galactic and extragalactic novae, however \citep[e.g.,][Healy et al. in prep.]{henze18, page22}.

Hachisu \& Kato \cite{hachisukato06} proposed a `universal decline law' for (classical) novae, considering optical and IR light-curves, which shows a similar shape to the `S'-class. Modelling the eruption as a radiatively-driven wind, their template has a decay slope steepening from $\alpha$~$\sim$~1.75 to $\sim$~3.5, around 6 mag below peak; this break is taken to be caused by an abrupt decrease in the wind mass-loss rate. As the wind ceases, the decline flattens to $\alpha$~$\sim$~3. However, \citep{strope10} find a much larger scatter in their measurements than can be physically explained by this universal model by \citep{hachisukato06}, and suggest this means there is an additional mechanism at work.
Indeed, \cite{shen22} show how the initial mass loss can be driven by binary interaction rather than radiation pressure.

\section{Summary}

Nova UV light-curves show a variety of shapes, evolutions and variabilities following the initial eruption. In some cases, the UV emission varies in phase (or anti-phase) with the X-rays, in others the wavebands seem completely unrelated (at least during the super-soft X-ray phase). In this paper, we concentrated on this latter population, where the UV decline can typically be modelled by a series of (usually) declining power-law slopes, while the X-rays first brighten through the SSS phase, and then fade. We have presented the first uniform analysis of nova UV light-curves -- including classical and recurrent, short-period and long -- which were well monitored by the {\em Neil Gehrels Swift Observatory}. Within this sample we find that all sources show a break in the UV decay slope around the point the SSS phase ends in the X-ray band, while additional breaks also occur over a range of times, both earlier and later.  While the light-curves using different filters for a given nova sometimes show very similar break times and slopes, at other times they disagree; there are no definitive trends between the different filter wavelengths in all cases.

Overall, there are no strong population trends between the evolution of the UV light-curves of the classical, symbiotic-like, short-period recurrent and long-period recurrent novae, although it is likely that the symbiotic-like, embedded systems will show a late-time break when the ejecta shock reaches the outer edge of the wind bubble formed by the giant companion. There is also no large difference in the range of unreddened, absolute magnitudes for the four populations considered, although the three symbiotic, long-period RNe are at the brighter end of the sample, while the majority of the CNe are significantly less luminous.


\vspace{6pt} 


\authorcontributions{Conceptualization, K.L.P., N.P.M.K., J.P.O.; methodology, K.L.P., N.P.M.K., J.P.O.; validation, K.L.P., N.P.M.K., J.P.O.; formal analysis, K.L.P., J.P.O.; visualization, K.L.P.; writing --- original draft preparation, K.L.P.; writing --- review and editing, K.L.P., N.P.M.K., J.P.O.
All authors have read and agreed to the published version of the manuscript.}

\funding{The {\em Swift} project in the UK is funded by the UK Space Agency. }

\dataavailability{The data underlying this paper are available in the Swift
archive at \\\url{https://www.swift.ac.uk/swift\_live/} and the HEASARC
Browse archive at \\\url{https://heasarc.gsfc.nasa.gov/cgi-bin/W3Browse/w3browse.pl}. 
}

\conflictsofinterest{The authors declare no conflict of interest.}

\begin{adjustwidth}{-\extralength}{0cm}

\reftitle{References}


\end{adjustwidth}
\end{document}